\documentclass[aip,jcp,reprint]{revtex4-1} 
\usepackage{graphicx}

\usepackage{amsmath}    
\usepackage{graphicx}   
\usepackage{color}
\usepackage{ulem}
\usepackage{caption}
\usepackage{subcaption}

\usepackage{array}
\usepackage{booktabs}
\usepackage{multirow}

\usepackage{booktabs}

\begin{document}
	\title{ Caliber based spectral gap optimization of order parameters (SGOOP) for sampling complex molecular systems}
	\author{Pratyush Tiwary} 
	 \affiliation{Department of Chemistry, Columbia University, New York 10027, USA.}

	\author{B. J. Berne}
	 \affiliation{Department of Chemistry, Columbia University, New York 10027, USA.}
	
	\date{\today}

	\begin{abstract}
	
In modern day simulations of many-body systems
much of the computational complexity is shifted to the
identification of slowly changing molecular order parameters
called collective variables (CV) or reaction coordinates. A vast
array of enhanced sampling methods are based on the
identification and biasing of these low-dimensional order
parameters, whose fluctuations are important in driving rare
events of interest. Here describe a new
algorithm for finding optimal low-dimensional collective variables for use in
enhanced sampling biasing methods like umbrella sampling,
metadynamics and related methods, when limited prior static
and dynamic information is known about the system, and a much larger
set of candidate CVs is specified. The algorithm involves
estimating the best combination of these candidate CVs, as
quantified by a maximum path entropy estimate of the spectral
gap for dynamics viewed as a function of that CV. Through multiple
practical examples, we show how this post-processing procedure
can lead to optimization of CV and several orders of magnitude
improvement in the convergence of the free energy calculated
through metadynamics, essentially giving the ability to extract
useful information even from unsuccessful metadynamics runs. 
	\end{abstract}
	\maketitle
	\section{Introduction}

With the advent of increasingly accurate force-fields and
powerful computers, Molecular Dynamics (MD) simulations have
become an ubiquitous tool for studying the static and dynamic
properties of systems across disciplines. However, most
realistic systems of interest are characterized by deep,
multiple free energy basins separated by high barriers. The
timescales associated with escaping such barriers can be
formidably high compared to what is accessible with
straightforward MD even with the most powerful computing
resources. Thus in order to accurately characterize such
landscapes with atomistic simulations, a large number of
enhanced sampling schemes have become popular, starting with the
pioneering works of Torrie, Valleau, Bennett and others
\cite{throwingropes,umbrella,bluemoon,hansmann1993prediction,voter_prl,meta_laio,wtm,abf,tamd,osrw,sisyphus2}.
 Many of these schemes involve probing the probability distribution along
selected low-dimensional collective variables (CVs), either
through a static pre-existing bias or through a bias constructed
on-the-fly, that enhances the sampling of hard to access but important regions in
the configuration space.
  
The quality, reliability, and usefulness of the
sampled distribution is in the end deeply dependent on the quality of
the chosen CV. Specifically, one key assumption inherent in
several enhanced sampling methods is that of time-scale separation \cite{szabo_timescale}:
for efficient and accurate sampling, the chosen CV should
encode all the relevant slow dynamics in the system, and any dynamics
not captured by the CV should be relatively fast. For most practical applications,
there are a large number of possible CVs that could be chosen, and it is not at
all obvious how to construct the best low-dimensional CV or CVs
for biasing from these various possible options. Success in enhanced
sampling simulations has traditionally relied on an apt use of
physical intuition to construct such low dimensional CVs.
Identification of good low dimensional CVs is in fact useful not just
for enhanced sampling simulations such as umbrella sampling and
metadynamics but also for distributed computing techniques like Markov
State Models (MSM) \cite{noe_jcp_2013}, allowing one to significantly
improve the quality and reliability of the constructed kinetic
models. Last but not the least, having an optimal low dimensional CV
can also help in the building of Brownian dynamics type models
\cite{bd_mccammon,morrone2011interplay}.  Indeed, given the importance of this
problem, there exists a range of methods that have been proposed to
solve it \cite{besthummer_rc,coifman2005,peters_rc,ma_dinner,diffusionmap,sketchmap,tuckerman_pnas2015,pvalue}.
  
In this communication, we report a new and computationally efficient
algorithm for designing good low-dimensional slow CVs. We suggest
that the best CV is one with the maximum separation of timescales
between visible slow and hidden fast processes \cite{coifman2008diffusion,szabo_timescale},
or the maximum spectral gap. The method is named  spectral 
gap optimization of order parameters (SGOOP).
Note that in this work henceforth we refer to the best CV in the singular,
without loss of any generality in the treatment. 
The notion of such a timescale separation is at the core of not just enhanced
sampling methods but also coarse-grained, Multiscale and projection
operator methods \cite{bernebook,cpmd,kevrekidis_equationfree}.  

Our algorithm involves learning the best linear or non-linear combination
of given candidate CVs, as quantified by a maximum path entropy \cite{caliber1}
estimate of the spectral gap for the dynamics of that CV. 
The input to the algorithm is any available information about the static and
dynamic properties of the system, accumulated through (i) a
biased simulation performed along a sub-optimal trial CV, possibly
(but not necessarily) complemented by (ii) short bursts of unbiased MD
runs, or (iii) by knowledge of experimental observables. Any type of
biased simulation could be used in (i), as long as it allows
projecting the stationary probability density estimate on generic CVs
without having to repeat the simulation. Metadynamics
\cite{tiwary_rewt} provides this functionality in a straightforward
manner and hence it is our method of choice here. 
Given this information we use  the principle of Maximum Caliber \cite{caliber1,caliber2}
 to set up an unbiased master equation for the dynamics of various trial
CVs. Through a simple post-processing optimization procedure we then find the CV 
with the maximal spectral gap of the associated transfer matrix. 
For instance, this optimization can be performed through 
a simulated annealing approach that maximizes the spectral gap by performing
a robust global search in the space of trial CVs. 

\begin{figure*}[t]
  \centering
        \includegraphics[height=2.3in]{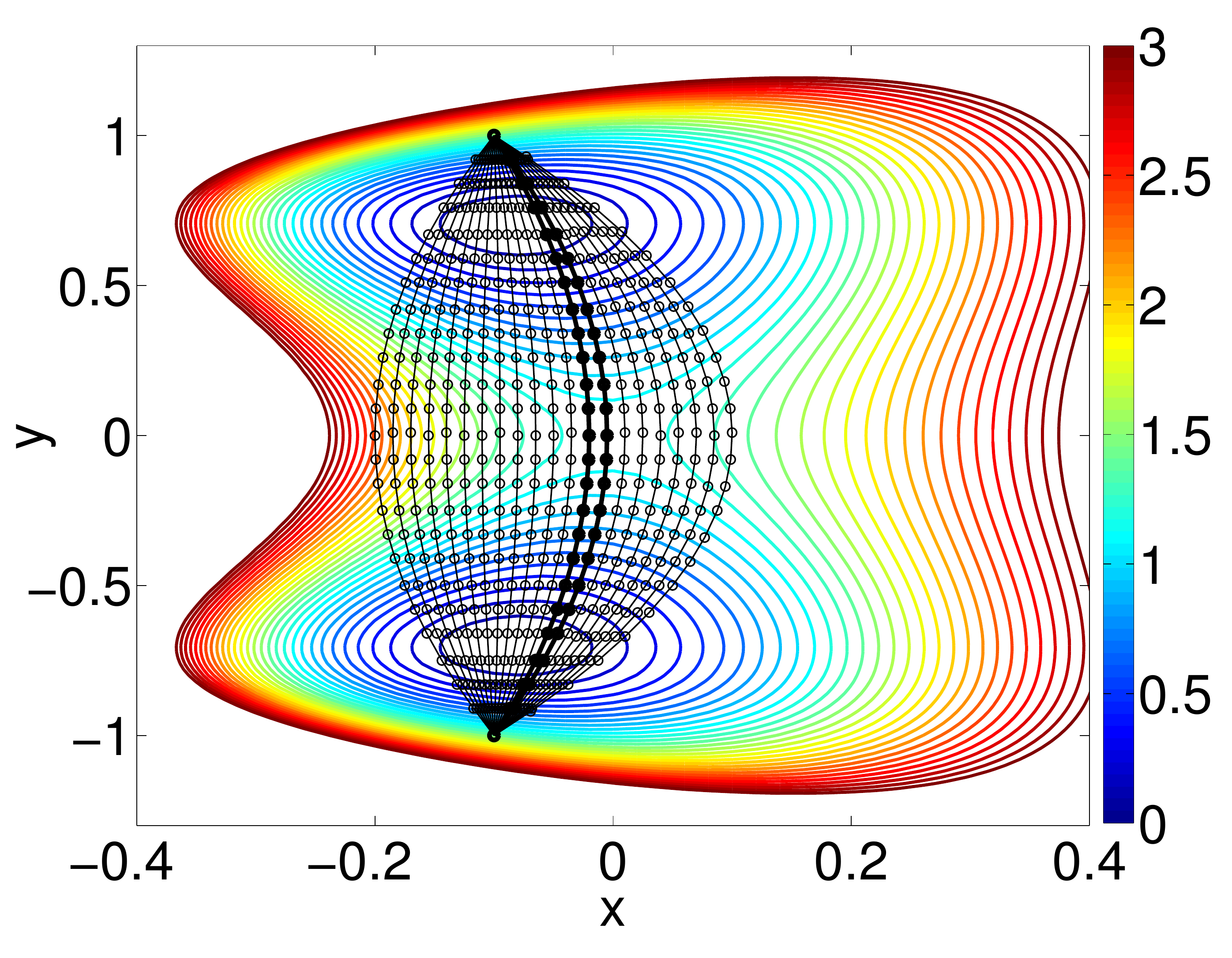} 
        ~
        \includegraphics[height=2.3in]{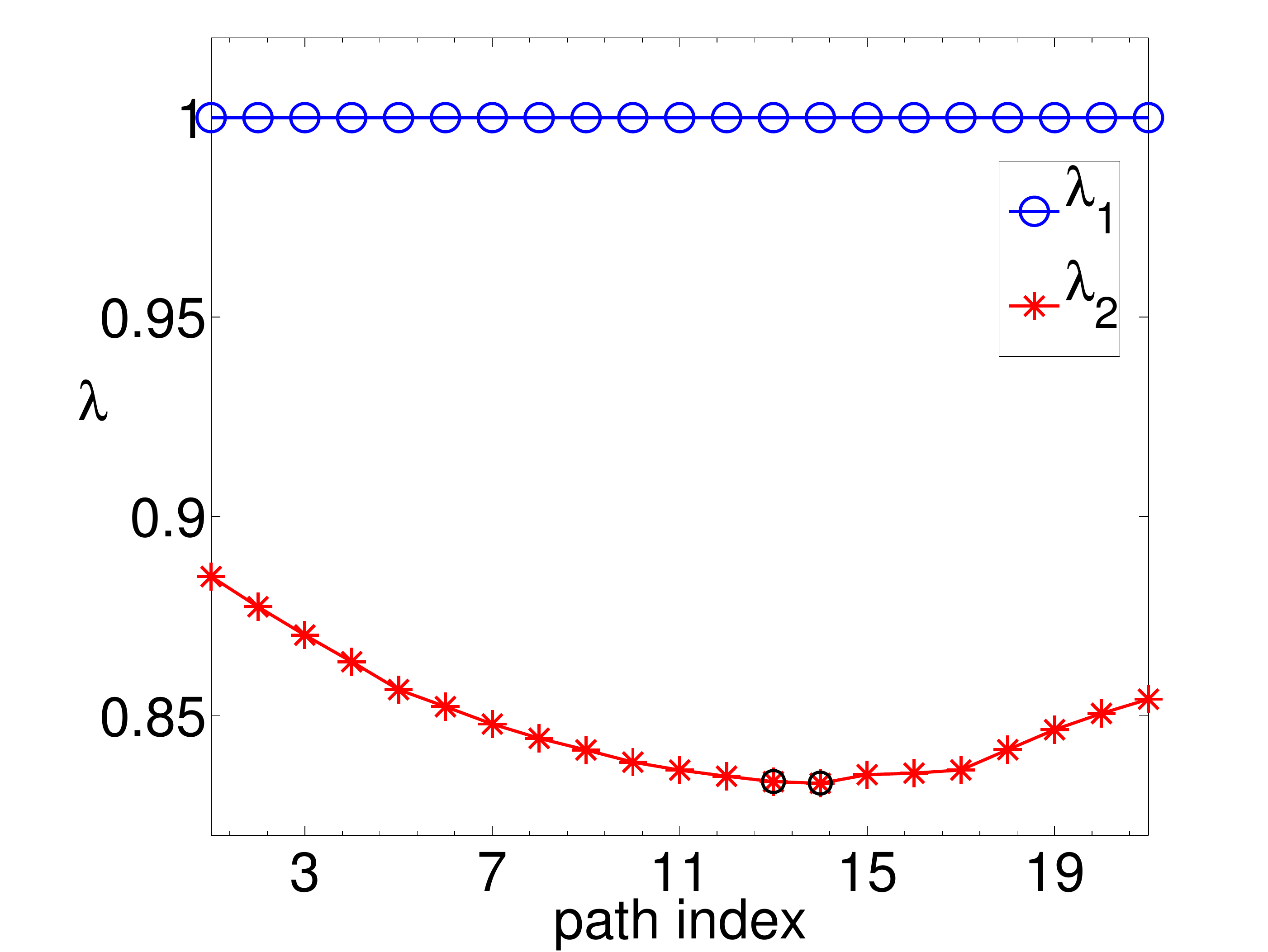}
\caption { In (a), we provide the 2-d De Leon-Berne potential  			
\cite{deleonberne} with several candidate path CVs imposed on it. Black 
circles denote the corresponding milestones \cite{fromatob}. See SI for further details of the CVs. 
In (b), the corresponding eigenvalues $\lambda_1$ and $\lambda_2$
 (i.e. excluding the stationary eigenvalue  $\lambda_0$) are 
shown for each of these paths.  As per the spectral gap
 given here by $\lambda_1 - \lambda_2$,  we identify two
 possible good paths marked with black circles in (b)
and correspondingly with thicker black lines in (a). Energy is in absolute units and $k_B T=0.1$.}
\label{db}
\end{figure*}

Through three practical examples, we show how our post-processing
procedure can lead to better choices of CVs, and to
several orders of magnitude improvement in the convergence of the free
energy calculated through the popular enhanced sampling technique
metadynamics. Furthermore, the algorithm is generally applicable irrespective of the 
number of stable basins. Our algorithm essentially provides the much needed
ability to extract useful information about relevant CVs even from
unsuccessful metadynamics runs. In addition to use in free energy
sampling methods, the optimized CV can
then also be used in other methods that provide kinetic rate
constants\cite{meta_time,fullerene}. We expect this algorithm to be of
widespread use in designing CVs for biasing during enhanced sampling
simulations, making the process significantly more automatic and far less reliant on
human intuition.

\section{Theory}
  	 
Let us consider a molecular system with $N$ atoms at temperature $T$. 
We assume there exists a large number $d$ of available order parameters with 
$1 \ll d \ll N$, collectively referred to as $\{\Theta \}$, such that the dynamics 
in this $d-$dimensional space is Markovian.
 These could be inter-molecular distances \cite{besthummer_rc}, torsional angles,
solvation states, nucleus size/shape \cite{tenfrenkel}, bond order parameters
\cite{steinhardt} etc.  The identification of such order parameters poses another 
complicated problem, but as routinely done in other methods aimed at optimizing CVs
\cite{besthummer_rc,noe_jcp_2013,tuckerman_pnas2015},
 we assume such order parameters are \textit{a priori} known.
			
There are several available biasing techniques that can sample the
 probability distribution of the space $\{\Theta \}$, and even calculate
  the rate constants for escape from stable states in this space \cite{meta_time}.  
 All of these techniques are feasible only for a very small number of CVs
  whose number is much smaller than $d$ -
typically one to three. These are the order parameters whose
fluctuations are deemed to be most important for the system or process
being studied, and by building a fixed or time-dependent bias of these CVs, 
one should be able to determine the true unbiased probability
distribution of the full space $\{\Theta \}$. But how does one decide what
is an optimal low-dimensional subset or combination of the available
order parameters?  This dimensionality reduction is of central importance to methods
such as umbrella sampling, metadynamics and others, the answer to
which decides the speed of convergence of the biased simulation, or if
it it will even ever converge within practically useful simulation
times.

\begin{figure*}[t]
\centering
\includegraphics[height=4.6in]{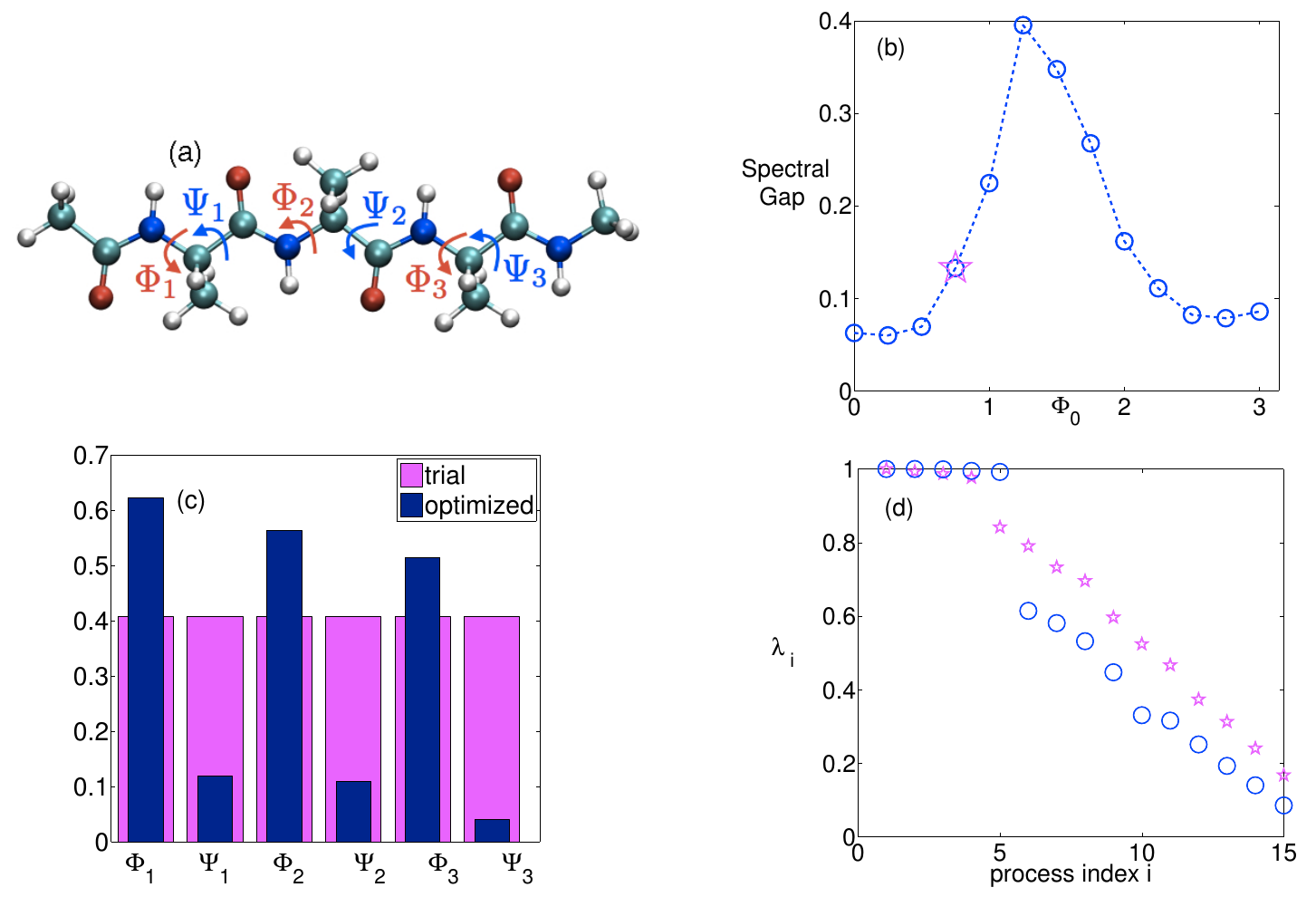}
\caption{(a) The 5-residue peptide studied in this work. The
  six dihedral angles are marked. (b) The output of the simulated
  annealing algorithm run separately for different $\theta_0$ values
  (blue circles). The starting value with the trial choice of CV is
  marked with a magenta colored star. (c) The trial (magenta) and optimized
  (blue) mixing coefficients $\{ c\}$ for the 6 dihedrals. (d) The spectrum of
  eigenvalues for dynamics projected on the trial (magenta) and
  optimized (blue) CVs. A distinct improvement can be seen in the spectral gap.}
\label{ala3}
\end{figure*}

The key idea in the current work is to perform enhanced sampling ({\it
  e.g.} metadynamics) with a choice of trial CVs,  complemented by information 
  gathered from short bursts of unbiased MD simulations and experimental
observables when available, to iteratively improve the CVs. The
maximum Caliber framework
\cite{jaynes_caliber,caliber1,caliber2,dixit2015inferring}, which is a
dynamical generalization of the hugely popular maximum entropy
framework \cite{jaynes1957}, provides a method for accomplishing
  this.

We start by choosing a trial CV given by  $f \{ \Theta \}$, where 
$f$ maps the space $ \{ \Theta \}$ onto a
lower dimensional space. The space along this trial
CV $f \{ \Theta \}$ is then  discretized in grids labeled $n$. This
CV could be multi-dimensional, with $n$ then indexing the
multidimensional grids. Let $p_n (t)$ denote the instantaneous
probability of the system being found in grid $n$. For the sake
of clarity, we assume that $f$ is a linear combination of $ \{
\Theta \}$, i.e. $f = c_1 \Theta_1 + c_2 \Theta_2 + ... + c_d
\Theta_d $. The treatment developed here applies to non-linear
combinations as well which we show in the examples. Then,
for a fixed $\Delta t$, we write a master equation:
\begin{align}
{\Delta p_n (t)\over \Delta t } = \Sigma_m \omega_{mn} p_m (t) -  \Sigma_m 		
\omega_{nm} p_n (t)  \equiv \Sigma_m \Omega_{nm} p_m (t) 
\label{eq:master}
\end{align}
where $\omega_{nm}$ is the rate of transition from grid $n$ to
$m$ per unit time. The matrix $\Omega_{nm} $ is the entirety of
all these rates. If the dynamics of $f \{ \Theta \}$ is
Markovian, then the matrix $\bf{k}$  of transition
probabilities is given for small $\Delta t$ by 
\begin{align}
{\bf k}= \exp({\bf{\Omega}} {{\Delta t}})\approx {\bf I} +
{\bf\Omega}{\Delta t}
\end{align}
  should not depend on the value of $\Delta t$
used in Eq. \ref{eq:master}. This provides a self-consistency
check of whether or not the CV so generated is Markovian. 
 In the maximum Caliber approach one uses all
available stationary state and dynamical information to construct
probabilities of micropaths. Instead of defining the entropy as a function of microstate
probabilities as in information theory and statistical
thermodynamics \cite{jaynes1957}, one now defines an entropy $S$ as a
functional of the probabilities of micropaths, which is essentially
a path integral. For the Markovian process of Eq.\ref{eq:master} \cite{filyukov}:
\begin{align}
S =  - \Sigma_{ab} p_a k_{ab} \text{ log } k_{ab}
\label{eq:caliber}
\end{align}
Path ensemble averages of time-dependent quantities $A_{ab}$ can
now be calculated as follows \cite{caliber1,caliber2}, where
the subscripts $a$,$b$ denote grid indices:
\begin{align}
\langle A \rangle =  \Sigma_{ab} p_a k_{ab} A_{ab}
\label{eq:pathaverage}
\end{align}			   
The path entropy of Eq. \ref{eq:caliber} incremented by
quantities accounting for constraints placed by our knowledge of
observables $\{ \langle A^n_{ab}  \rangle \}$, and some other constraints such as
detailed balance, is collectively called Caliber
\cite{caliber1,caliber2}. Maximizing the Caliber is then equivalent to being
least non-committal about missing dynamic and static information, with the end result
being that one obtains a relation between the grid-to-grid rates
and the stationary probabilities as follows:
\begin{align}
\omega_{ab} = {1 \over \Delta t}{\sqrt{   p_b \over p_a     }}e^{-\Sigma_i \rho_i A_{ab}^i}
\label{eq:localdynamics}
\end{align}
Here $i$ runs over the number of available dynamical pieces of
information, and $\rho_i$ is the Lagrange multiplier for the
associated constraint. As a special case, consider when the only
observable at hand is the mean number of transitions in observation interval $\Delta
t$ over the $entire$ grid \cite{caliber2} along a trial CV. In this case,
the above equation takes a particularly simple
and useful form:
\begin{align}
\omega_{ab} = {1 \over \Delta t}{\sqrt{   p_b \over p_a     }}e^{- \rho}
\label{eq:onlyentire}
\end{align}			
Our method then involves calculating for various trial CVs the
spectral gap of the transition probability matrix \textbf{k}, which
for $a \neq b$ is $k_{ab} = \omega_{ab} \Delta t$ and
satisfies normalization $\Sigma_b k_{ab} = 1$.  Let
$\{\lambda\}$ denote the set of eigenvalues of \textbf{k}, with
$\lambda_0 \equiv 1 > \lambda_1 \geq \lambda_2 ...$. The spectral gap
is then defined as $\lambda_s - \lambda_{s+1} $, where $s$ is the
number of barriers apparent from the free energy estimate projected on
the CV at hand, that are higher than a user-defined threshold
(typically $\gtrsim k_B T$).  Estimating the Lagrange
multiplier is computationally expensive, so a first estimate for
maximizing the spectral gap is performed using Eq.  \ref{eq:onlyentire} where 
the Lagrange multiplier $\rho$ need not be computed. 
Also note that in the limit of small $\Delta t$, the matrix $\textbf{k}$
will be diagonally dominated \cite{dham}, and to estimate the spectral gap
one needs only an accurate estimate of relative local free energies.
More static or dynamical information
\cite{szabo_bicout,hummer2005position,laio_kinetics,berne1968jaynes,nmrmeta,metainference}
simply introduces additional Lagrange multipliers
 and can be treated through Eq. \ref{eq:localdynamics}.
This can be done if the intention is to calculate an accurate kinetic
model with correct estimates of the dominant eigenvalues and not just
the spectral gap.

We are now in a position to describe the actual algorithm. It
comprises the following two steps in a sequential manner, and
can be improved by iterating.
\begin{enumerate} 
\item Perform
metadynamics along a trial CV $f = c_1 \Theta_1 + c_2 \Theta_2 +
... + c_d \Theta_d $ to get a crude estimate of the stationary
density. 
\item  As post-processing, perform  optimization  in
  the space of mixing coefficients $\{c_1,c_2... c_d \}$ to identify
  the CV with the maximal spectral gap. The reweighting functionality
  \cite{tiwary_rewt} of metadynamics allows projection of free energy
  estimates on different CVs with minimal computational effort, and is
  used to calculate the \textbf{k} matrix through Eq. \ref{eq:onlyentire}.
  We elaborate on the optimization procedure details
  in the next section (Illustrative Examples).
\end{enumerate}

The optimization procedure gives the best CV as the one with highest
spectral gap, given the information at hand. As in any maximum entropy
framework \cite{jaynes1957}, the better the quality of this
information, the more accurate will be the spectral gap. But even with
very poor quality information, as we show in the examples, the
algorithm still leads to significant improvements in the CV. 
Furthermore, whether or not the CV is Markovian
 can also be checked by repeating step 2 for different time intervals $\Delta t$ of observation and
determining if the spectral gap is independent of the value of $\Delta t$.
	
	\begin{figure*}[t]
\centering
		        \includegraphics[height=2.5in]{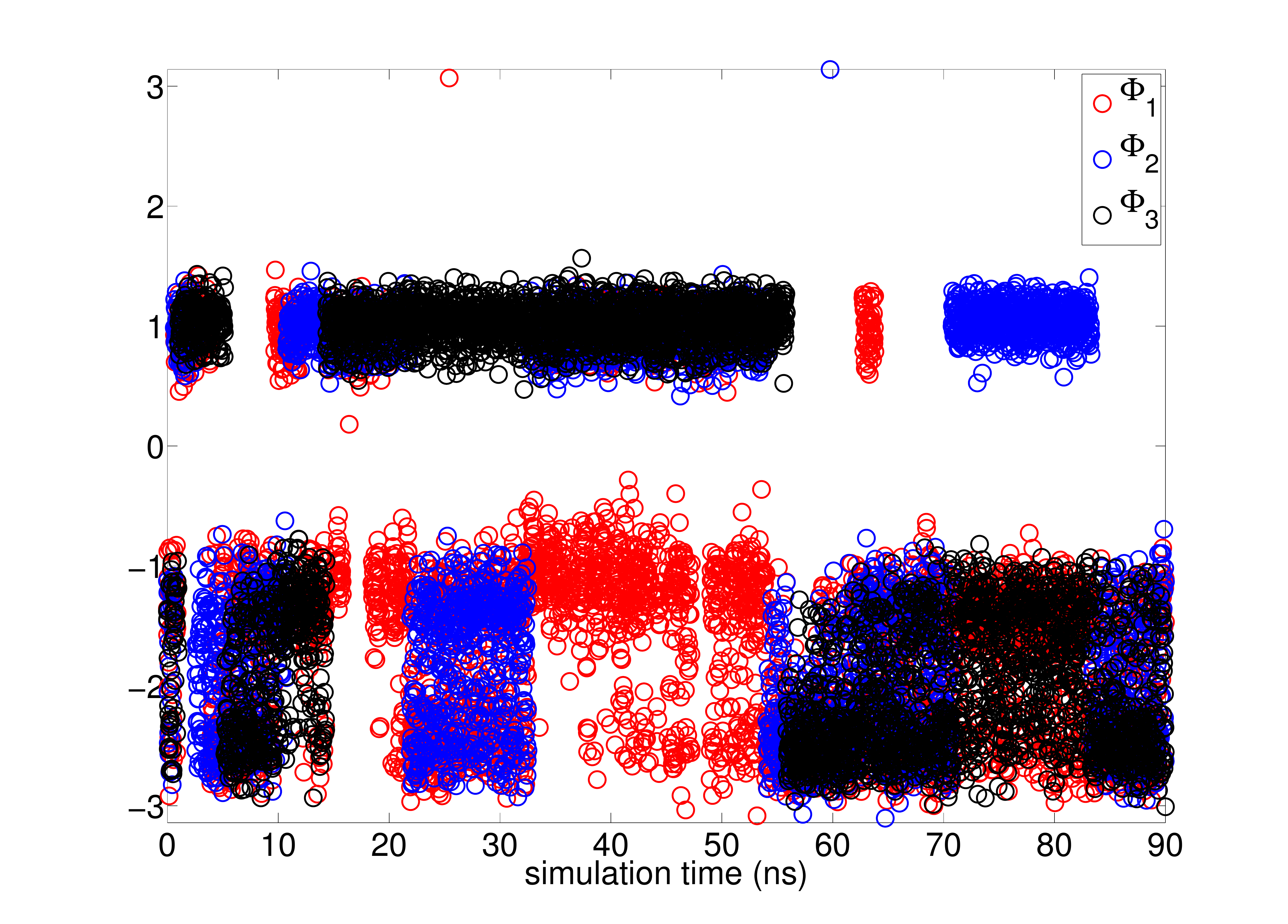}
		        \includegraphics[height=2.5in]{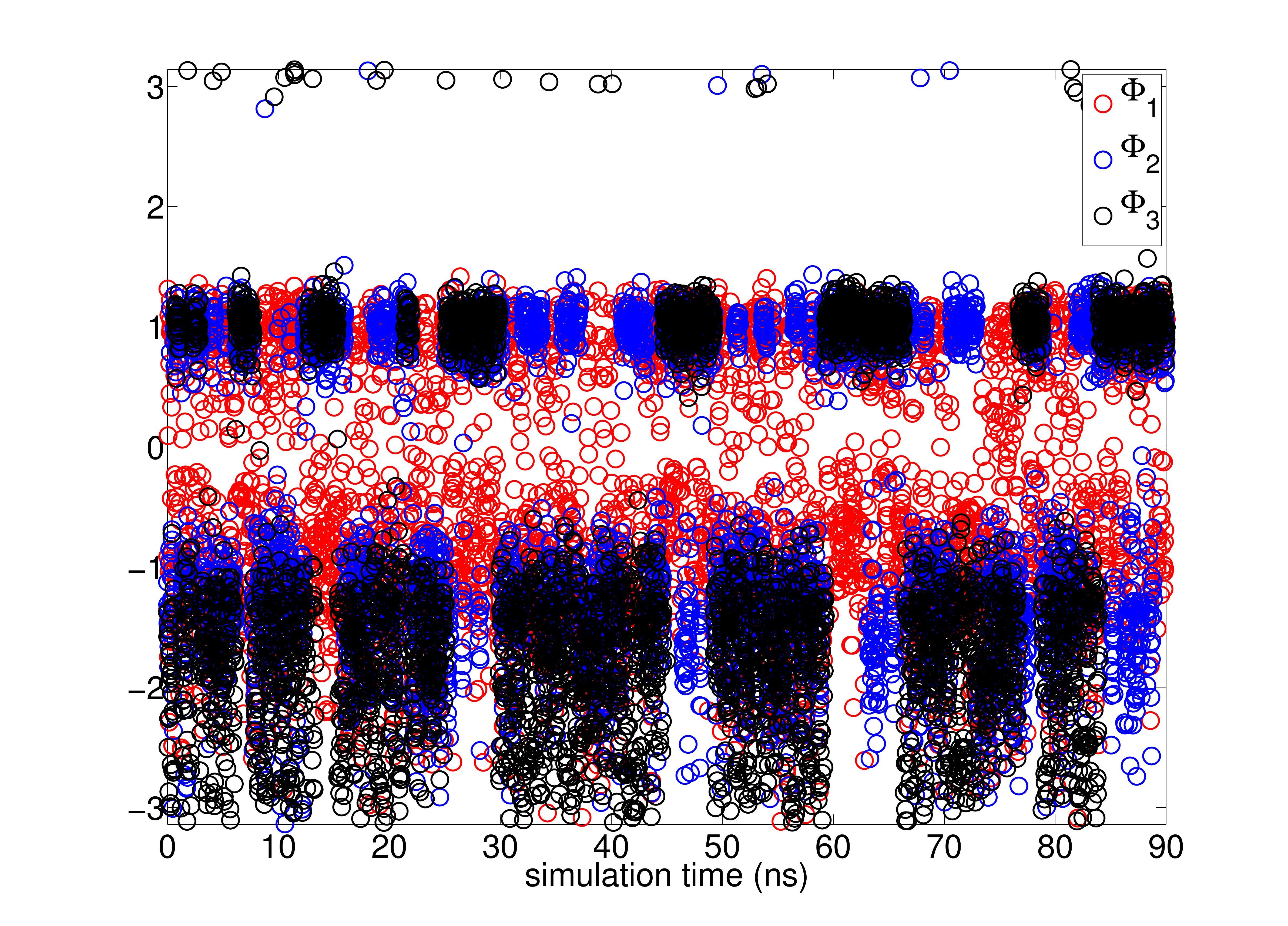}
		\caption { (a) and (b) show trajectories obtained from metadynamics biasing the 
		trial CV and the optimized CV respectively. First 20 ns of the trajectory shown in (a) was used to generate the optimized CV for (b). A very pronounced improvement in
		 the enhancement of sampling can be seen with the optimized CV. }
		\label{traj}
\end{figure*}

\section{\large{Illustrative Examples}}
\subsection{Model 2-d landscapes: The De Leon-Berne potential}
The first illustrative example for SGOOP is a model 2-state potential introduced
by De Leon and Berne \cite{deleonberne}. To sample this landscape at
temperature $k_B T=0.1$, we perform metadynamics with path CVs, a class of
widely used CVs that can capture non-local and non-linear fluctuations
(see \cite{fromatob} for details). Path CVs require specification of a series of 
milestones between two points in configuration space, where the milestones can be
described in terms of generic order parameters. Fluctuations in the system
 can then be enhanced in the direction along and perpendicular to these milestones, 
 leading to efficient exploration of the space.  In Fig. \ref{db} (a) we show the
2-d potential along with several possible path CVs imposed on it. We first perform a
short trial metadynamics run biasing the y-coordinate. By post-processing this,
we generate the spectral gaps for various paths using
Eq. \ref{eq:onlyentire} (Fig. \ref{db} (b)). By comparing
Fig. \ref{db}(a) against Fig. \ref{db}(b), it is clear how the path
with maximum spectral gap is the minimum energy pathway
passing through the saddle point. In this case while this result could
have simply been obtained through Nudged Elastic Band type
calculations \cite{neb} - the point is to use this example to develop
intuition for the method. Also note that moving around the best path
to others that are a bit distant from it, does not lead to much
change in the spectral gap. This is consistent with the observation
that in several enhanced sampling methods such as metadynamics or
umbrella sampling \cite{umbrella, meta_laio,wtm}, the CV need not be
precisely the true reaction coordinate, as long as it has a
sufficient overlap with it \cite{fromatob,trypsin}. 

In the Supplemental Information (SI),
we provide a similar analysis on another 2-d model potential but with 3 states. 
The conclusions are similar.

\subsection{5-residue peptide}
Now we move to a more complex system, which has also been considered
as a test case for new enhanced sampling methods
\cite{omar_variational} in order to establish their usefulness. 
This is the 5-residue peptide $Ace-Ala_3
-Nme$ in vacuum (see Fig. \ref{ala3} (a)), where there are six possibly relevant
dihedral torsion angles. Here we ask the question: what is the best
possible 1-d linear combination of these six dihedrals that we could
bias but still maximally enhance exploration of the 6-d space
comprising all the dihedrals?

					    \begin{figure*}[t]
		    \begin{subfigure}[t]{0.3\textwidth}
		        \includegraphics[height=1.95in]{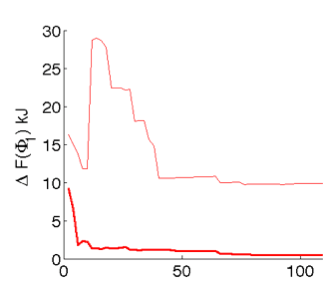}

		    \end{subfigure}%
		    ~ 
		    \begin{subfigure}[t]{0.3\textwidth}
		        \includegraphics[height=1.9in]{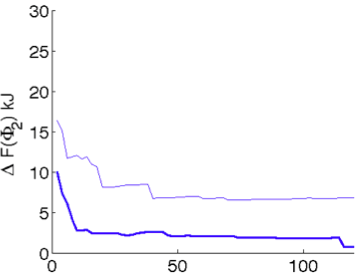}

		       \end{subfigure}
			    ~ 
		    \begin{subfigure}[t]{0.3\textwidth}
		        \includegraphics[height=1.9in]{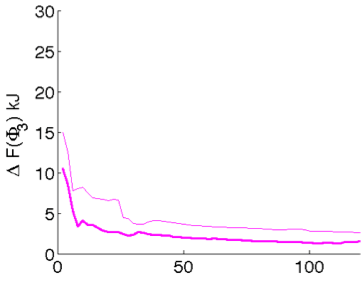}

		       \end{subfigure}
                      \caption { Errors in the 1-d free energies in kJ calculated with
		    respect to reference free energies \cite{omar_variational,omar_jctc}	
  		    using the error metric from \cite{adaptive}. Thin and thick lines
                     denote values using the trial and optimized CVs respectively. }
		    \label{fes}
		    \end{figure*}

In this problem, for periodicity related numerical reasons, we bias a reference cosine
defined by cos($\theta - \theta_0$), where $\theta$ is one of the six
dihedral angles, and $\theta_0$ is some reference value whose optimal
choice we do not know \textit{a priori}.  Through our algorithm we
then seek to identify:
\begin{enumerate}
\item[(a)] The best choice of mixing coefficients $\{ c \}$ to use in trial CV
  $f = c_1 \Phi_1' + c_2 \Psi_1' + c_3 \Phi_2' + c_4 \Psi_2' + c_5
  \Phi_3' + c_6 \Psi_3' $, where we keep the euclidean norm of $\{ c
  \}$=1, and for any angle $\theta$ the prime denotes the transformation
  $\theta \mapsto 0.5 + cos(\theta - \theta_0)$.
\item[(b)] The best choice of $\theta_0$, kept same for all 6 dihedrals. 
\end{enumerate}
		
We start with the trial CV where all members of $\{ c \}$ are the same
subject to euclidean norm of $\{ c\}$=1, and an arbitrary choice of
$\theta_0 = 0.75$ radians is taken. A short metadynamics run is
performed biasing this trial CV.  See supplemental information (SI)
for details of the metadynamics and MD parameters \cite{plumed2}, and Fig. \ref{traj}
(a) for the metadynamics trajectory used for spectral gap optimization.  Based on the free energy
estimate generated from this run, a simulated annealing procedure is
performed in the space $\{ c\}$ for various $\theta_0$ values. Starting from the spectral gap
estimated using Eq. \ref{eq:onlyentire} for the trial CV, this involves executing Metropolis
moves in the $\{ c\}$  space with an attempt to find the global maxima of
the spectral gap. In Fig. \ref{ala3} (b-d) respectively, we show how the
spectral gap is increased by the simulated annealing procedure, and
the corresponding best estimate of $\{ c,\theta_0\}$. The algorithm
suggests the minimal role of the angles $\Psi_1,\Psi_2,\Psi_3$ as can
be seen through their relatively low weights \cite{omar_variational} (Fig. \ref{ala3}
(c)). The spectrum of eigenvalues for dynamics projected on the trial
(magenta) and optimized (blue) CVs, along with respective spectral
gaps is provided in Fig. \ref{ala3}(d). Fig. \ref{traj} (a-b) show the
metadynamics trajectories for the three dihedral angles
$\Phi_1,\Phi_2,\Phi_3$ with the trial and the optimized CVs
respectively. A very pronounced improvement in the quality of sampling can be
seen. Fig. \ref{fes} (a-c) shows the rate of convergence of the error
of the estimated free energy \cite{tiwary_rewt} with respect to
reference values from other approaches \cite{omar_variational},
through metadynamics runs performed with each of
the trial and optimized CVs respectively. The error metric is the same
as in \cite{adaptive, omar_variational}, and is calculated for all
points within 25 kJ of the global minimum in the respective 1-d free
energy. The behavior is robust with respect to the choice of this
threshold value. As can be seen, the optimized CV, even though it was
obtained on the basis of a very poorly converged and short (20 ns)
metadynamics run, leads to several orders of magnitudes improvement in
the rate at which the free energies converge. Interestingly,
iterating the algorithm with the improved 1-d CV did not lead to much
improvement in the sampling, reflecting that the optimized
coefficients $\{ c\}$ are close to the best that can be achieved with
a 1-d CV for this problem.


\section{\large{Conclusions}}
To conclude, we have introduced a new approach named SGOOP (spectral gap optimization of order parameters) for improving the
choice of low-dimensional CVs for  biasing in
enhanced sampling in complex systems. This is accomplished
through the use of maximum Caliber based spectral gap estimates.  The
algorithm is iterative in spirit, and attempts to learn how to
improve CVs  based on available stationary and
dynamic data. We also provide several proof-of-concept practical
examples to establish the potential usefulness of the method.  For
 model 2-d potentials the algorithm was shown to yield the
minimum energy pathway. For a small peptide, we found very significant
improvement in determining the best 1-d collective variable 
from six possible functions with no \textit{ad hoc} or \textit{intuition based} tuning. 
Future work will use this algorithm to treat a range of problems,
especially involving protein-ligand unbinding. For instance, 
the displacement of water molecules and protein flexibility 
are often slowly varying order parameters in
unbinding \cite{trypsin,justaddwater,berneweekszhou,fullerene}, but do
we really need to bias one or both of these for the purpose of
sampling?  Another issue to be considered in
future work is can we use these
optimized CVs to obtain reliable dynamical information from
metadynamics \cite{meta_time,pvalue}, including the
  very important off-rate for ligand unbinding \cite{copeland2006drug,trypsin}. 
  
  One central
limitation of this algorithm is having to specify
possibly a large number of order parameters
  that may be important. But for many physical problems one does
have a sense of which order parameters could be at work, and this is
where we expect this algorithm to be of tremendous use. Another obvious
 limitation is with systems devoid of a time scale separation \cite{zwanzig1990rate}
  - for example, in glassy systems where there is an effectively continuous 
  spectrum of eigenvalues with no discernible time scale separation.
However, the dynamics of many complex and real-world molecular systems does thankfully show a time scale separation between few relevant slow modes and remaining fast ones \cite{sethna_science}, and we
expect our algorithm to be of help in unraveling the thermodynamics and dynamics
in such systems.

\begin{acknowledgments}
We  would like to thank Purushottam Dixit for helpful discussions
regarding Caliber, Omar Valsson for providing system set-up and
reference free energies for the peptide, and Jed Brown for originally
suggesting a spectral gap approach.  This work was supported by
 grants from the National Institutes of Health [NIH-GM4330] and the 
Extreme Science and Engineering Discovery Environment (XSEDE) [TG-MCA08X002].
\end{acknowledgments}

\end{document}